\documentstyle[amssymb,aps,pre,twocolumn, floats, epsfig]{revtex}

\newcommand{\myhsize}{2in}
\newcommand{\myhSize}{2.2in}
\newcommand{\myvspace}{1mm}

\begin{document}
\title{{\Large {A new class of semiclassical wave function uniformizations}}}
\author{Ji\v{r}\'{\i} Van\'{\i}\v{c}ek and Eric J. Heller}
\date{\today}
\address{Department of Physics, Harvard University}
\address{Cambridge, MA 02138}
\maketitle

\begin{abstract}
We present a new semiclassical technique which relies on replacing
complicated classical manifold structure with simpler manifolds, which
are then evaluated by the usual semiclassical rules. Under
circumstances where the original manifold structure gives poor or
useless results semiclassically the replacement manifolds can yield
remarkable accuracy. We give several working examples to illustrate
the theory presented here.
\end{abstract}

\section{Introduction}

Semiclassical methods are based on classical mechanics; the relevant
classical manifolds form the ``skeleton'' to which the wave function is
attached. However, associating a semiclassical wave function with a classical
manifold is problematic when enclosed loops in phase space have area less
than $\hbar $. If that happens, typically two or more stationary-phase
points corresponding to distinct contributions to the semiclassical wave
function are not well separated in phase. Consequently, the stationary-phase (SP) method, on which semiclassical approximation is based, breaks down. Worse,
in some important situations, such as the universal homoclinic oscillations
associated with chaotic regions of phase space, the bad enclosed loops may
be repeated many times in a small region\cite{softChaos}.

Although the semiclassical wave function diverges near caustics lurking near
the small area loops, the exact quantum wave function is of course
well-behaved. This is one manifestation of quantum smoothing over classical
detail. We shall quantify below how this smoothing occurs. We suggest and
test a new way of looking at the smoothing process in terms of ``replacement
manifolds'', in which new, well behaved classical manifolds are substituted
for the original badly behaved ones. One application of this approach is
uniformizing a semiclassical wave function in the vicinity of the ubiquitous
homoclinic oscillations of a chaotic system.

It is widely understood that very small changes in potentials or walls of
billiards can have little effect on quantum eigenstates, but can drastically
affect the classical manifold structure. This immediately implies a
many-to-one relationship between the classical manifolds and quantum wave
functions. That is, many different underlying classical manifold patterns
correspond to the same wave function. Now, whatever the choice of manifold
(amongst these equivalent forms) we can suppose an appropriate
uniformization exists which gives the correct wave function. However, the
non-uniformized, simple semiclassical limit for the various choices of
manifolds can differ drastically in accuracy. It is this fact which we
exploit in the present paper.

\section{Standard uniformization: The Airy function}

\label{standardUnif}
The paradigm example of uniformization is the Airy function, i.e. the energy
eigenfunction of the linear ramp potential.
We take the potential to be of the form 
\begin{equation}
V(x)=-\beta x
\end{equation}
The Schr\"{o}dinger equation is 
\begin{equation}
{-{\frac{\hbar ^{2}}{2m}}}\psi _{E}^{^{\prime \prime }}(x)-(\beta x+E)\psi
_{E}(x)=0
\end{equation}
After the shift 
\begin{equation}
\psi _{E}(x)=\psi _{0}(x+E/\beta )
\end{equation}
and scaling transformation 
\begin{equation}
u={\frac{m^{1/3}\beta ^{1/3}}{\hbar ^{2/3}}}x\equiv \gamma x
\end{equation}
the Schr\"{o}dinger equation reads 
\begin{equation}
{{\frac{1}{2}}}\Psi ^{^{\prime \prime }}(u)+u\Psi (u)=0;  \label{airy}
\end{equation}
with $\Psi (u)=\psi _{0}(x)$. Eq.~(\ref{airy}) is exactly solved by going to
the momentum representation. We have 
\begin{equation}
\left( {{\frac{1}{2}}}p^{2}-i{\frac{\partial }{\partial p}}\right) \Phi (p)=0
\end{equation}
with solution 
\begin{equation}
\Phi (p)=e^{-ip^{3}/6}
\end{equation}
Thus 
\begin{eqnarray}
\Psi (u) &=&{\frac{1}{\sqrt{2\pi }}}\int\limits_{-\infty }^{\infty }e^{ip\
u}\Phi (p)\ dp  \nonumber \\
&=&{\frac{2}{\sqrt{2\pi }}}\int\limits_{0}^{\infty }\cos ({p^{3}/6-p\ u})\ dp
\nonumber \\
&=&2^{1/2}{\rm Ai}\left( -2^{1/3}u\right)
\end{eqnarray}
and 
\[
\psi (x)=\eta \ {\rm Ai}(-2^{1/3}\gamma x)=\eta \ {\rm Ai}\left( {\frac{%
-2^{1/3}m^{1/3}\beta ^{1/3}}{\hbar ^{2/3}}}x\right) 
\]
SP approximation for the momentum integral gives 
\begin{eqnarray}
\Psi (u) &=&\left[ {\frac{2\pi }{-i(-\sqrt{2u})}}\right] ^{1/2}e^{-(\sqrt{2u}%
)^{3}/6+i\sqrt{2u}\ u}+{\rm c.c.}  \nonumber \\
&\sim &{\frac{1}{u^{1/4}}}\cos \left({\frac{2}{3}}\sqrt{2}u^{3/2}-{\frac{\pi }{4}}%
\right)
\end{eqnarray}
Noting that 
\begin{equation}
\int^{u}p(u^{\prime })\ du^{\prime }={\frac{(2u)^{3/2}}{3}}
\end{equation}
we get finally 
\[
\Psi (u)\sim {\frac{1}{\sqrt{p(u)}}}\cos \left(\int^{u}p(u^{\prime })\ du^{\prime
}-{\frac{\pi }{4}}\right) 
\]
This is the standard WKB form for the linear ramp potential. 
\begin{figure}[htbp]

\centerline{\epsfig{figure=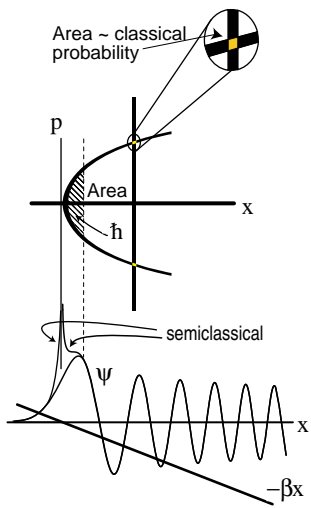,width=\myhSize}}
\vspace{\myvspace}
\caption{Diagram showing the  areas in phase space important to the
  semiclassical approximation to the Airy function.} 
\label{areaai}

\end{figure}
Figure \ref{areaai} illustrates the situation for the semiclassical position
wave function, which becomes inaccurate when the area enclosed by the
vertical $x$-line and the energy contour corresponding to the state $\vert
E\rangle$ falls below Planck's constant. Upon entering the ``bad'' region,
the original momentum integral may be substituted for the WKB result, giving
a uniformization. We have an integral over ``initial momentum,'' making this
example a type of ``initial value representation'' (IVR) as first discussed
by Miller\cite{ivr,millerIVR}. Of course in this case we know the result of the
integral is an Airy function, but in general there may be integrals which
are not known analytically or even close to ones that are known.

In this paper we encounter more complex situations which are induced by
nonlinear interactions, including chaos, reflection from corrugated
surfaces, etc. We will show that intricate manifolds can sometimes be replaced
by much simpler ones, which may themselves be evaluated semiclassically in
any basis, using a much simplified uniformization if required.

\section{Replacement manifolds}

\begin{figure}[htbp]

\centerline{\epsfig{figure=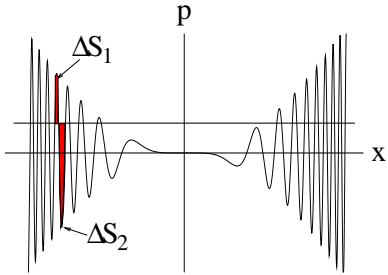,width=\myhsize}}
\vspace{\myvspace}
\caption{Areas in phase space important to momentum wave function
  associated with the manifold $p(x)=p_{0}-{\frac{3}{ 
2}}\hbar \alpha \lambda x^{2}\sin (\alpha x^{3})$, $\lambda =0.5$. Primitive
semiclassical approximation fails since both areas $\Delta S_{1}$ and
$\Delta S_{2}$ can be smaller than $\hbar $. }
\label{oldfig1}

\end{figure}

We now introduce an example of such a  problematic Lagrangian
manifold, defined by a position-momentum relation 
\begin{equation}
p(x)=p_{0}-\frac{3}{2}\,\hbar \lambda \alpha x^{2}\sin \alpha x^{3}.
\label{MAN1}
\end{equation}
This model was first introduced in Ref.\cite{rick} but was not explored
there. The function (\ref{MAN1}) has oscillations of increasing amplitude
and frequency, which however have the same area $\int p(x)dx=\lambda \hbar $
between successive zeros (see Fig.~\ref{oldfig1},  small value of $\lambda$ corresponds to small area
of loops). This is also a
characteristic property of homoclinic oscillations near a periodic orbit.
The momentum $p(x)$ spans an ever larger range as $x$ increases, but we now
show that for $\lambda <1$ the manifold's semiclassical behavior can be
understood by replacing it with three smooth manifolds. The rules for
replacement are simple to derive. 

\begin{figure}[htbp]

\centerline{\epsfig{figure=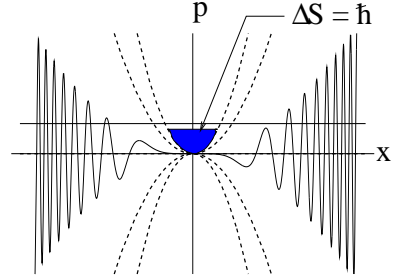,width=\myhsize}}
\vspace{\myvspace}
\caption{Original (solid line) and replacement manifolds (dashed lines) for $%
p(x)=p_{0}-{\frac{3}{2}}\hbar \alpha \lambda x^{2}\sin (\alpha x^{3})$, $%
\lambda =0.5$.}
\label{oldfig2}

\end{figure}

The action function $S(x)$ is the integral of $p(x)$: 
\begin{equation}
S(x)=\int_{0}^{x}p(x^{\prime })\,dx^{\prime }=p_{0}x+\frac{1}{2}\,\hbar
\lambda \left( \cos \alpha x^{3}-1\right) .  \label{ACT1}
\end{equation}
The ``wave function'' 
\begin{equation}
\psi (x)=A(x)\,e^{iS(x)/\hbar }  \label{WF1}
\end{equation}
can be approximated by 
\begin{equation}
\psi (x)\approx A(x)\,e^{-i\lambda /2}e^{ip_{0}x/\hbar }\left( 1+i\frac{%
\lambda }{2}\cos (\alpha x^{3})\right)
\end{equation}
\begin{eqnarray}
\psi (x) &\approx &A(x)\,e^{-i\lambda /2}e^{ip_{0}x/\hbar } \\
&&\times \left( 1+i\frac{\lambda }{4}e^{i\alpha x^{3}}+i\frac{\lambda }{4}%
e^{-i\alpha x^{3}}\right)  \nonumber
\end{eqnarray}
if $\lambda <1$. For small $\lambda $ this expression is nearly identical
with (\ref{WF1}) with the original action (\ref{ACT1}), because $\cos
(\alpha x^{3})$ is bounded by $\pm 1$. However semiclassically it has the
interpretation of the sum of three smooth classical manifolds, i.e. $%
p=p_{0}$, $p=p_{0}+3\,\hbar \alpha x^{2}$ and $p=p_{0}-3\,\hbar \alpha
x^{2}$. They have the weights $A(x)$, $i\frac{\lambda }{4}e^{-i\lambda
/2}A(x)$, $i\frac{\lambda }{4}e^{-i\lambda /2}A(x)$, respectively. The
situation is depicted in Fig.~\ref{oldfig2}.

Almost every discussion of the relation of classical and quantum mechanics
for chaotic systems alludes to quantum smoothing, but here we have seen
explicitly one way this smoothing comes about. For all reasonable
purposes the three smooth classical manifolds accurately
replace the rapid oscillations of the original manifold. Note too that
depending on the parameters the outlying manifolds $p=p_{0}+3\,\hbar \alpha
x^{2},\;\;$ and $p=p_{0}-3\,\hbar \alpha x^{2}$ lie far beyond the limits of
the original distributions. We shall discuss interesting consequences of
this in the next section.

So far, we have considered the wave function only in position representation where
caustics are absent and the semiclassical form (\ref{WF1}) was accurate. The
situation deteriorates drastically in momentum space. In order to find
the  wave function in momentum representation, we have to sum over all contributions from
intersections of a horizontal line (corresponding to a momentum eigenstate)
with the oscillating manifold (\ref{MAN1}). When $\lambda <1$, the adjacent
intersections will be separated by a phase $\Delta S\,/\,\hbar$ that is smaller than 1 for any
classically allowed momentum, and therefore the standard semiclassical approximation
will break down for {\it all classical momenta}! Usual Airy-type
uniformization methods of Section \ref{standardUnif} (also see
\cite{airyCarrier,airyBerry}) do not serve under these circumstances,
because the manifold is dense with caustics (see
the filled-in areas $\Delta S_{1}$ and $\Delta S_{2}$ in Fig.~\ref{oldfig1}%
).  Apparently, we have no other choice but to
part with semiclassical method and use a numerical Fourier transform (IVR)
over the position wave function to obtain $\phi (p)$. However using
the replacement manifolds (RMs) is a much simpler and more intuitive approach
which can rescue the situation. 
To find the momentum wave function $\phi (p)$ associated with the
original manifold we add (with appropriate weights) wave
functions $\phi _{n}(p)$ corresponding to the three RMs.
Each of these partial wave functions is equal to a Fourier transform of
a corresponding position wave function $\psi _{n}(x)$, 
\begin{equation}
\phi _{n}(p)=\frac{1}{\sqrt{2\pi \hbar }}\int dx\,\psi
_{n}(x)\,e^{-ipx/\hbar }.
\end{equation}
Except for a small region close to $p_{0}$ (highlighted in Fig.~\ref
{oldfig2}), this integral can be evaluated by SP method! In
other words, although the semiclassical approximation completely fails for
the original manifold, it works almost everywhere for the replacement
manifolds. Moreover, as we show in the following section, the standard Airy
uniformization procedure of Section \ref{standardUnif} can be exploited to correct the inaccuracy in a narrow region near $p_{0}$. We will see that the momentum wave function
found by applying SP approximation to RMs is in
excellent agreement with numerical solution. We emphasize that the
RMs do not yield the original, badly behaved semiclassical
result, but rather something much more accurate and much simpler. 

\section{Uniformization with replacement manifolds in a generalized model}

\label{GENERAL}

In order to capture other properties of homoclinic oscillations in our
model, we consider a generalized manifold 
\begin{equation}
p(x)=p_{0}-\frac{1}{2}\hbar \lambda \frac{d\xi (x)}{dx}\sin \xi (x),
\label{MAN}
\end{equation}
in which $\alpha x^{3}$ from (\ref{MAN1}) is replaced by a generic smooth
function $\xi (x)$. The presence of the derivative $\frac{d\xi (x)}{dx}$
ensures equal loop areas between successive zeros of $\xi (x)$. Example of a manifold
with $\xi (x)=\alpha \log \frac{x}{a}$
is displayed in Fig.~\ref{rep_man_for_log}.
Manifold (\ref{MAN}) can be obtained from a horizontal manifold representing a
momentum state $|p_{0}\rangle $ by a canonical transformation of
coordinates\cite{canonical} 
\begin{equation}
x=x_{0},\;p=p_{0}-\frac{1}{2}\hbar \lambda \frac{d\xi }{dx_{0}}\sin \xi
(x_{0}),
\end{equation}
generated by a function
\begin{equation}
F_{2}(x,p_{0})=p_{0}x+\frac{1}{2}\hbar \lambda \cos \xi (x),
\end{equation}

We find the uniform semiclassical transformation element $\langle
p|p_{0}\rangle $, and thereby fix the amplitude prefactor to be $A(x)=(2\pi
\hbar )^{-1/2}$; this however retains all the problematic features of the
manifold. We solve the problem to all orders in $\lambda $ by replacing the
original manifold with an infinite series of RMs. 
First we note that the semiclassical transformation element 
\begin{equation}
\langle x|p_{0}\rangle _{sc}=\frac{1}{\sqrt{%
2\pi \hbar }}\,e^{iF_{2}(x,p_{0})/\hbar }
\end{equation}
is fairly accurate since there are no caustics in this representation (to
each final $x$ corresponds a single initial $x_{0}$). The
momentum-representation element is obtained by a Fourier transform, 
\begin{eqnarray}
\label{IVR_integral}
\langle p|p_{0}\rangle &\simeq &\frac{1}{\sqrt{2\pi \hbar }}\int
dx\,e^{-ipx/\hbar }\,\langle x|p_{0}\rangle _{sc}  \label{IVRform} \\
&=&\frac{1}{2\pi \hbar }\int dx\,\exp \left\{ \frac{i}{\hbar }\left[
-px+F_{2}(x,p_{0})\right] \right\} .  \nonumber
\end{eqnarray}
If evaluated by SP method, the integral yields a very poor semiclassical
result since there exist many coalescing SP points. An accurate uniform
answer is obtained by evaluating integral (\ref{IVR_integral}) exactly. The accuracy is
usually further improved by changing the integration variable from $x$ to $%
x_{0}$ (in our case both forms are equivalent since $x=x_{0}$). This
uniform version of $\langle p|p_{0}\rangle $ is in IVR form\cite{ivr,millerIVR},
but can best be evaluated by writing $\langle x|p_{0}\rangle $ as a sum over
RMs. Recognizing that the factor $e^{i\frac{\lambda }{2}%
\cos \xi (x)}$ in $\langle x|p_{0}\rangle _{sc}=(2\pi \hbar )^{-1/2}\exp
\left[ ip_{0}x/\hbar +i\frac{\lambda }{2}\cos \xi (x)\right] $ is a
generating function for Bessel functions, we can extend the sum from
previous section beyond the first order in $\lambda $. In fact we obtain an
infinite sum convergent for any $\lambda $, 
\begin{eqnarray}
\langle x|p_{0}\rangle &=&\frac{1}{\sqrt{2\pi \hbar }}e^{ip_{0}x/\hbar
}\sum_{n=-\infty }^{\infty }J_{n}\left( \frac{\lambda }{2}\right)
i^{n}e^{in\xi (x)} \\
&=&\sum_{n=-\infty }^{\infty }J_{n}\left( \frac{\lambda }{2}\right)
i^{n}\langle x|p_{0}\rangle _{n}  \nonumber
\end{eqnarray}
where $\langle x|p_{0}\rangle _{n}=(2\pi \hbar )^{-1/2}\exp \left\{ i\left[
p_{0}x+n\hbar \xi (x)\right] /\hbar \right\} $. We can rewrite (\ref{IVRform}) 
as 
\begin{eqnarray}
\langle p|p_{0}\rangle &=&\sum_{n=-\infty }^{\infty }J_{n}\left( \frac{%
\lambda }{2}\right) \frac{i^{n}}{\sqrt{2\pi \hbar }}\int dx\,e^{-ipx/\hbar
}\,\langle x|p_{0}\rangle _{n}  \label{RM_SUM} \\
&=&\sum_{n=-\infty }^{\infty }J_{n}\left( \frac{\lambda }{2}\right)
i^{n}\langle p|p_{0}\rangle _{n}  \nonumber
\end{eqnarray}
where 
\begin{equation}
\langle p|p_{0}\rangle _{n}=\frac{1}{2\pi \hbar }\int dx\,\exp \left\{ \frac{%
i}{\hbar }\left[ (p_{0}-p)x+n\hbar \xi (x)\right] \right\}  \label{PROP}
\end{equation}
may be identified semiclassically as a transformation element corresponding
to the $n$-th RM 
\begin{equation}
p_{n}(x)=p_{0}+n\hbar \frac{d\xi (x)}{dx},
\end{equation}
This manifold for most functions $\xi (x)$ of interest contains no caustics,
and consequently $\langle p|p_{0}\rangle _{n}$ allows evaluation by
SP method! As promised, we have expressed the uniform version
of $\langle p|p_{0}\rangle $ as a weighted sum over semiclassical
replacement manifolds,
\[
\langle p|p_{0}\rangle _{unif}=\sum_{n=-\infty
}^{\infty }J_{n}\left( \frac{\lambda}{2}\right) i^{n}\langle p|p_{0}\rangle
_{n,sc}.
\]%
If applied to a well-behaved (i.e. decaying fast enough at $\pm \infty $%
) function of $p_{0}$, the resulting sum converges for any $\lambda $.
The replacement-manifold method is therefore not restricted to the regime
where loop areas are smaller than $\hbar $. It also works in the strongly
chaotic regime where $\lambda >1$ (and where the standard semiclassical
approximation holds) if all manifolds up to $|n|\simeq \lambda $ are
included in the sum. This follows since $J_{n}\left( \lambda /2\right) $
considered as a function of $n$ decays exponentially fast for $|n|>\lambda
/2 $. In other words, we only need to include several RMs
beyond those intersecting the original manifold. Physically, the manifolds
outside of the range of the original one lie in a classically forbidden
region in which the wave function is exponentially suppressed. 

In opposite case ($\lambda <1$) which interests us the most, considering RMs up
to $|n|=1$ will suffice while the ``primitive'' semiclassical approximation fails
completely even for $p=p_{0}$.
In this regime $\lambda$ can be thought of as a parameter defining the strength
of a perturbation which causes the manifold to oscillate around a
simple manifold that describes the system if perturbation is absent. 

\begin{figure}[htbp]

\centerline{\epsfig{figure=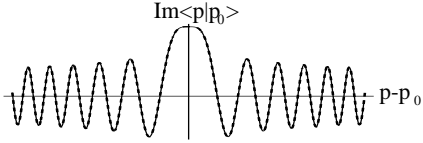,width=\myhSize}}
\vspace{\myvspace}
\caption{Momentum wave function: comparison of numerical IVR (solid line) with
the RM sum up to $|n|=1$ (O($\lambda $), dashed line) for original manifold with $\xi (x)=\alpha x^{3}$, $\lambda =0.5$ and $\alpha = 1$.}
\label{wf_for_ax3}

\end{figure}

To complete our solution for a specific function $\xi (x)$, we must evaluate 
$\langle p|p_{0}\rangle _{n}$. For $n=0$, we cannot use SP approximation
(because the action is linear), but integral~(\ref{PROP}) is trivial and we
obtain in general 
\begin{equation}
\langle p|p_{0}\rangle _{0}=\delta (p-p_{0}).
\end{equation}
For $n\neq 0$, SP method applied to (\ref{PROP}) yields a result in the form
of a sum over SP points $x_{sp}$, 
\begin{eqnarray}
&&\langle p|p_{0}\rangle _{n,sc} = \frac{1}{\sqrt{2\pi \hbar }}
\label{SP_INT} \\
&&\times \sum_{x=x_{sp}}\left| \frac{\partial ^{2}f_{n}}{\partial x^{2}}%
\right| ^{-1/2}\exp \left[ i\left( \frac{f_{n}}{\hbar} +\frac{\pi }{4}%
\mathop{\rm sgn}%
\frac{\partial ^{2}f_{n}}{\partial x^{2}}\right) \right] ^{{}}  \nonumber
\end{eqnarray}
where $f_{n}(x,p,p_{0})=(p_{0}-p)\,x+n\hbar \xi (x)$ and $\frac{\partial
f_{n}}{\partial x}=0$ for $x=x_{sp}$. For manifold (\ref{MAN1}),
each RM $p_{n}(x)=p_{0}+3n\hbar \alpha x^{2}$ has two SP points 
\begin{equation}
x_{sp}=\pm \left( \frac{p-p_{0}}{3n\hbar \alpha }\right) ^{1/2}
\end{equation}
whose contributions add to give 
\begin{eqnarray}
\langle p|p_{0}\rangle _{n,sc} &=&\pi ^{-1/2}[3\,n\hbar ^{3}\alpha
(p-p_{0})]^{-1/4} \\
&& \times \cos \left[ \frac{2}{(n\alpha )^{1/2}}\left( \frac{p-p_{0}}{3\hbar 
}\right) ^{3/2}-\frac{\pi }{4}\right] .  \nonumber
\end{eqnarray}
RMs still contain one caustic at $p=p_{0}$, which can be easily uniformized
by evaluating integral (\ref{PROP}) exactly\cite{airyUnif}, 
\[
\langle p|p_{0}\rangle _{n,unif}=\frac{1}{(3\,|n|\alpha )^{1/3}\hbar }%
Ai\left( -\frac{(%
\mathop{\rm sgn}%
\,n)(p-p_{0})}{(3|n|\alpha )^{1/3}\hbar }\right) . 
\]
Having found $\langle p|p_{0}\rangle _{n}$, we can calculate $\langle
p|p_{0}\rangle $ from (\ref{RM_SUM}) to all orders in $\lambda $. 
$\langle p|p_{0}\rangle _{1,sc}$ and $\langle p|p_{0}\rangle _{1,unif}$
agree very well except for a small region near $p_{0}$ (similarly as in Fig.~
\ref{areaai}). Fig.~\ref{wf_for_ax3} demonstrates the excellent agreement of
RM method with the direct numerical computation of $\langle p|p_{0}\rangle $
using fast Fourier transform (IVR). 

Here
the replacement-manifold approach should be compared to the
semiclassical  perturbation (SCP)
approximation of Miller and Smith \cite{millerSPT,hubbard} who used
perturbative classical dynamics to calculate the semiclassical $S$ matrix.
 They first applied this
method to the collinear collision of an atom with a diatomic molecule \cite{millerSPT}. The
starting point is the initial value representation of the classical
scattering matrix,
\begin{eqnarray}
S_{n_{f},n_{i}} &=& \frac{1}{2\pi }\int_{0}^{2\pi }dq_{i}\left[ \frac{\partial
q_{f}(q_{i},n_{i})}{\partial q_{i}}\right] ^{1/2} \\
\nonumber
&\times & \exp \left( i\left\{ \Phi
(q_{i},n_{i})+q_{f}(q_{i},n_{i})\left[ n_{f}(q_{i},n_{i})-n_{f}\right]
\right\} \right) 
\end{eqnarray}
where $\hbar =1$, $n_{i}$, $n_{f}$, $q_{i}$, $q_{f}$ denote, respectively,
the initial and final values of the action and angle variables describing
the internal degree of freedom, and $\Phi $ is the action integral. For
details, see Ref. \cite{ivr} and \cite{millerSPT}. If the classical quantities listed above are
calculated using the first-order perturbation dynamics, the $S$ matrix takes
the form
\[
S_{n_{f},n_{i}}=\frac{e^{i\Phi _{0}}}{2\pi }\int_{0}^{2\pi }dq_{i}\exp
\left\{ -i\left[ \left( n_{f}-n_{i}\right) q_{i}+A(q_{i},n_{i})\right]
\right\} 
\]
where $\Phi _{0}$ is twice the phase-shift for the unperturbed potential and 
$A(q_{i},n_{i})$ is the time integral of the perturbation potential along
the unperturbed trajectory \cite{millerSPT}.  For various physical systems (such as
ion-dipole collisions \cite{millerSPT} or atom-surface scattering \cite{hubbard}), $%
A(q_{i},n_{i})$ has a sinusoidal dependence on $q_{i}$ as long as
perturbation remains small. In this regime we expect the SCP and RM
approximations to give the same answer.  We shall verify
that for the scattering by a corrugated wall in Section \ref{WALL}. The SCP approximation of Miller and Smith presently appears to
be more widely applicable since it does not require any special property of
the manifold.  On the other hand, advantage of the RM method lies in
the fact that it is not subjected to the smallness of perturbation. Provided
that the \textit{full} action falls into one of the classes discussed above,
we can apply the RM method without approximating classical dynamics.

\section{Model of homoclinic oscillations}

\begin{figure}[htbp]

\centerline{\epsfig{figure=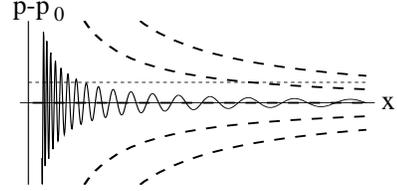,width=\myhsize}}
\vspace{\myvspace}
\caption{Original (solid line) and replacement manifolds (dashed lines) for $%
\xi (x)=\alpha \log {\frac{x}{a}}$, $\lambda =0.5$.}
\label{rep_man_for_log}

\end{figure}

\begin{figure}[htbp]

\centerline{\epsfig{figure=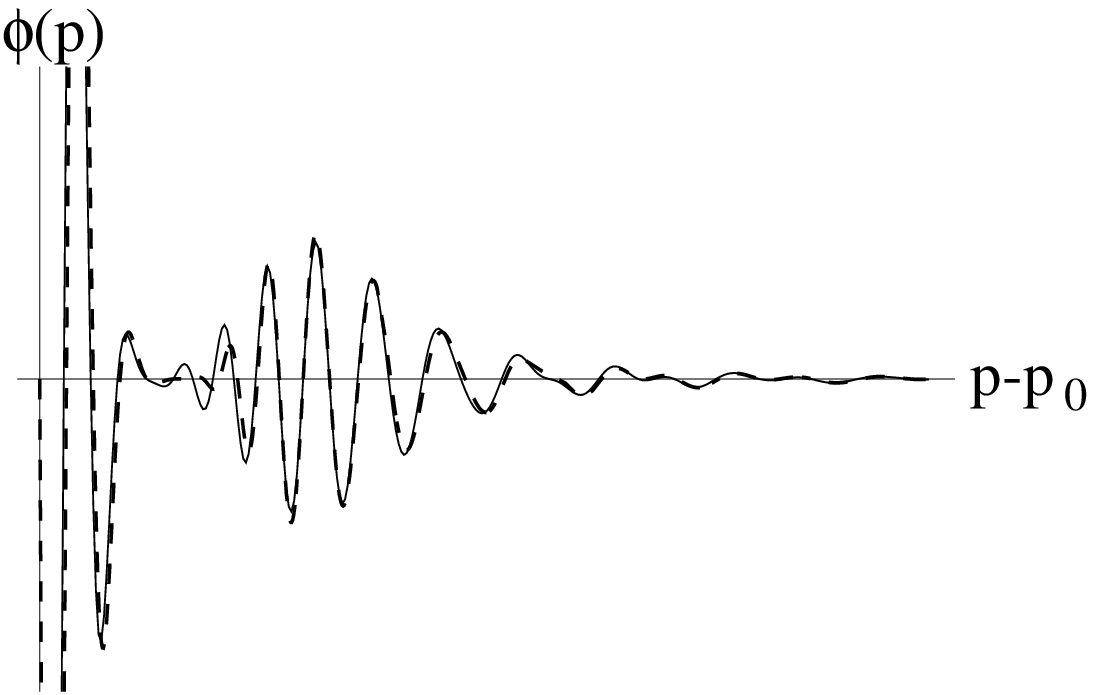,width=\myhsize}}
\vspace{\myvspace}
\caption{Momentum wave function: comparison of numerical IVR (solid line) with the RM sum up to $j=1$ ($O(\lambda)$, dashed line) for original manifold with $\xi(x) = \alpha \log
\left( \frac{x}{a}\right)$, $\lambda=0.5$ and $\alpha \approx 33$.}
\label{wfForLog}

\end{figure}

Homoclinic oscillations in chaotic systems have another characteristic
property: the amplitude of oscillations (i.e. ``height'' of loops) increases
exponentially as we approach an unstable periodic orbit (see Fig.~\ref
{rep_man_for_log}). 
To keep area of loops constant, the width of loops has
to decrease accordingly. Because systems in nature are usually bounded, the
exponentially growing loops must eventually twist in a complicated way to
fit into available phase space. For sake of simplicity, we consider a model
that satisfies the requirement of exponential growth, but is unbounded. Such
a model can be found among generalized manifolds of previous section.
Specifically, we want to find a manifold whose homoclinic points (i.e.
intersections with line $p=p_{0}$ in our case) form a geometric series, 
\begin{equation}
x_{n}=a\,e^{n\lambda _{+}}
\end{equation}
where $\lambda _{+}$ is the Lyapunov exponent. In other words,
we are looking for a function $\xi (x)$ such that $\xi (x_{n})=2\pi n$. The
logarithmic function comes to mind first since $\log \frac{x_{n}}{a}%
=n\lambda _{+}$. With the correct prefactor $\alpha =\frac{2\pi }{\lambda
_{+}}$, we find the desired function 
\begin{equation}
\xi (x)=\alpha \log \frac{x}{a}  \label{XI2}
\end{equation}
because $\xi (x_{n})=\alpha \log \frac{x_{n}}{a}=\frac{2\pi }{\lambda _{+}}%
n\lambda _{+}=2\pi n$. Substituting $\xi (x)$ from (\ref{XI2}) into the
general form (\ref{MAN}), we obtain a manifold described by 
\begin{equation}
p=p_{0}-\frac{1}{2}\frac{\hbar \lambda \alpha }{x}\cos (\alpha \log \frac{x}{%
a})  \label{MAN2}
\end{equation}
and displayed in Fig.~\ref{rep_man_for_log}. Replacement manifolds (19)
become 
\begin{equation}
p_{n}(x)=p_{0}+\frac{n\hbar \alpha }{x}.
\end{equation}
In the present case, function $f_{n}$ from (\ref{SP_INT}) becomes
\[
f_{n}(x,p,p_{0})=(p_{0}-p)x+n\hbar \alpha \log \frac{x}{a},
\]
and its single
SP point, 
\begin{equation}
x_{sp}=\frac{n\hbar \alpha }{p-p_{0}}.
\end{equation}
For $p>p_{0}$, only manifolds with $n>0$ contribute, and for $p<p_{0}$ only $%
n<0$ is allowed. Defining $j=|n|$, for $p\neq p_{0}$ we replace the sum over 
$n$ by sum over all positive $j$. In both cases, the $j$-th manifold gives a
semiclassical contribution 
\begin{eqnarray}
\langle p|p_{0}\rangle _{j,sc}&=&\left( \frac{j\alpha }{2\pi }\right) ^{1/2}%
\frac{1}{|p-p_{0}|}  \label{PROP2SC} \\
\times && \exp \left\{ i\,%
\mathop{\rm sgn}%
(p-p_{0})\left[ j\alpha \log \frac{j\hbar \alpha }{ae|p-p_{0}|}-\frac{\pi }{4%
}\right] \right\}  \nonumber
\end{eqnarray}
with $j=|n|>0$. In this case caustics are missing because there is a single
SP point $x_{sp}$. We have a finite integration limit at zero, but it is
separated from $x_{sp}$ by an infinite action (equal to the area delimited
by lines $x=0$, $p=p_{0}$, and the $n$-th RM, see Fig.~\ref{rep_man_for_log}%
). In fact, it can be shown that further terms in asymptotic expansion of $%
\langle p|p_{0}\rangle _{j,sc}$ have the same dependence on $\hbar $ and $%
p-p_{0}$ as (\ref{PROP2SC}), and only differ in their dependence on $j\alpha 
$. Luckily, these claims can be easily verified by evaluating integral (\ref
{PROP}) analytically. This oscillatory integral is made convergent by
displacing momentum $p$ with an imaginary infinitesimal term $-i\epsilon $, $%
\epsilon >0$, 
\begin{eqnarray}
&&\langle p|p_{0}\rangle _{n,unif}=\frac{1}{2\pi \hbar } \label{INT2}\\
&\times & \int_{0}^{\infty }dx\,\exp \left\{ \frac{i}{\hbar }\left[
-(p-i\epsilon -p_{0})x+n\hbar \alpha \log \frac{x}{a}\right] \right\}
. \nonumber
\end{eqnarray}
In this form, the answer is found by rotating the contour about the origin
of complex plane by $-\frac{\pi }{2}%
%TCIMACRO{\func{sgn}}
%BeginExpansion
\mathop{\rm sgn}%
%EndExpansion
(p-p_{0})$. 
\begin{figure}[htbp]

\centerline{\epsfig{figure=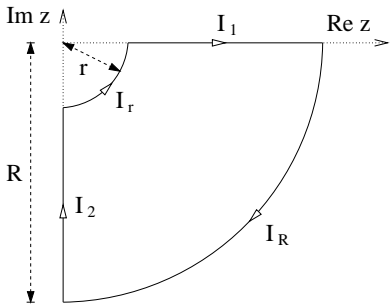,width=\myhsize}}
\vspace{\myvspace}
\caption{Contour for evaluation of $\langle p|p_{0}\rangle _{n,unif}$ in the
case $\xi (x)=\alpha \log {x/a}$.}
\label{contour}

\end{figure}
Let us explicitly solve the case $p>p_{0}$. Fig.~\ref{contour} shows a
contour in complex plane enclosing a region with no singularities, and as a
result the sum of integrals along appropriate parts of the contour is zero, $%
I_{1}+I_{R}+I_{2}+I_{r}=0$. In the limit $R\rightarrow \infty $, $%
r\rightarrow 0$, $I_{1}$ becomes our desired integral (\ref{INT2}) and both $%
I_{R}$ and $I_{r}$ vanish, implying $I_{1}=-I_{2}$. Contour for integral $%
I_{2}$ is the set of complex points $z=-ix=e^{-i\pi /2}x$, $x>0$.
Consequently, 
\begin{eqnarray}
&&-I_{2}=\frac{1}{2\pi \hbar }e^{-i\pi /2} \times \\
&&\int_{0}^{\infty }\!\!\!dx\,\exp \left\{ \frac{i}{\hbar }[-(p-p_{0}-i\epsilon
)(-ix)+n\hbar \alpha \log \frac{xe^{-i\pi /2}}{a}]\right\}.  \nonumber
\end{eqnarray}
Transforming to a dimensionless variable $y=\frac{(p-p_{0}-i\epsilon )x}{%
\hbar }$, the integral becomes 
\begin{eqnarray}
-I_{2} &=&\frac{1}{2\pi \hbar }\exp \left\{ (n\alpha -i)\frac{\pi }{2}%
-in\alpha \log \left[ (p-p_{0}-i\epsilon )a/\hbar \right] \right\}  \nonumber
\\
&\times& \frac{\hbar }{p-p_{0}-i\epsilon } \int_{0}^{\infty
}dy\,e^{-y}y^{in\alpha }.  \nonumber
\end{eqnarray}
Recognizing the remaining integral as a gamma function of complex argument $%
1+in\alpha $, 
\begin{eqnarray}
&&-I_{2}=\frac{\Gamma (1+in\alpha )}{2\pi (p-p_{0}-i\epsilon )}  \nonumber \\
&&\times \exp \left\{ (n\alpha -i)\frac{\pi }{2}-in\alpha \log \left[
(p-p_{0}-i\epsilon )a/\hbar \right] \right\} .
\end{eqnarray}
For $p<p_{0}$, we need to close the contour in the upper half-plane, but
otherwise the procedure is analogous. Combining the two cases, we obtain a
single analytic expression for the uniform transformation element, 
\begin{eqnarray}
\langle p|p_{0}&&\rangle _{n,unif} =\frac{\Gamma (1+in\alpha )}{2\pi
(|p-p_{0}|-is\epsilon )} \\
\times &&\exp \left[ (n\alpha -i)\frac{\pi }{2}s-in\alpha \log \frac{%
(|p-p_{0}|-is\epsilon )a}{\hbar }\right]  \nonumber
\end{eqnarray}
where $s=%
%TCIMACRO{\func{sgn}}
%BeginExpansion
\mathop{\rm sgn}%
%EndExpansion
(p-p_{0})$. As promised, the uniform and semiclassical forms have
identical dependence on $\hbar $ and $p-p_{0}$. The only difference lies in
the dependence on $n\alpha $, in particular the semiclassical form is
obtained if we keep only the first term in the asymptotic expansion of $%
\Gamma (1+in\alpha )$ in $\langle p|p_{0}\rangle _{n,unif}$.
However, for our conservative choice of $\alpha \approx 33$ the agreement between $%
\langle p|p_{0}\rangle _{n,sc}$ and $\langle p|p_{0}\rangle _{n,unif}$ is
such that they may be used interchangeably for any practical purposes.
Finally, in Fig.~\ref{wfForLog}, $\langle p|p_{0}\rangle $ is applied
to an initial Gaussian wave packet centered around $X_{0}$, $P_{0}$, namely
\[
\phi ^{i}(p_{0})=\left( \frac{\pi \hbar ^{2}}{\sigma ^{2}}\right)
^{-1/4}\!\!\!\!\!\!\!\exp \left[ \frac{i}{\hbar }(P_{0}-p_{0})X_{0}-\frac{\sigma
^{2}(p_{0}-P_{0})^{2}}{2\hbar ^{2}}\right] .
\]
The final position wave function is
\[
\psi _{sc}^{f}(x)=(\pi \sigma ^{2})^{-1/4} \exp \left[ \frac{i}{\hbar }%
F_{2}(x,P_{0})-\frac{(x-X_{0})^{2}}{2\sigma ^{2}}\right] 
\]
and the uniform momentum wave function,
\begin{eqnarray}
&&\phi _{unif}^{f}(p)=J_0\left(\frac{\lambda}{2}\right) \phi ^{i}(p) \\
&+&\sum_{j=1}^{\infty } J_j\left(\frac{\lambda}{2}\right) i^{\lambda} \langle p|p_{0}\rangle
_{j,sc}\frac{(2\pi \hbar )^{1/2}}{(\pi \sigma ^{2})^{1/4}}\exp \left[ -\frac{%
(x_{sp,j}-x_{0})^{2}}{2\sigma ^{2}}\right] \nonumber
\end{eqnarray}
with $x_{sp,j}=j\hbar \alpha /|p-p_{0}|$ and $\langle p|p_{0}\rangle _{j,sc}$
given by (\ref{PROP2SC}). The large oscillation near $p=p_0$  in Fig.~\ref{wfForLog} corresponds to the zeroth RM\
contribution $\phi ^{i}(p)$, the smaller wavelet to the right corresponds to  the remaining RMs.  The figure confirms the excellent accord
between RM method and numerical IVR evaluation by FFT\cite{complexSP}.   We have thus succeeded in uniformizing something with many of the properties of homoclinic
oscillations near an unstable periodic orbit using replacement-manifold
``technology''.

\section{Comparison with the stationary-phase method}

It is instructive to check explicitly if the expansion in terms of
RMs agrees with semiclassical method applied to the
original manifold when areas of loops are larger than $\hbar $ ($\lambda >1$%
). Let us choose an analytically solvable example with $\xi (x)=\alpha x$.
Corresponding manifold 
\begin{equation}
p(x)=p_{0}-\frac{1}{2}\,\hbar \lambda \alpha \sin \alpha x  \label{MAN3}
\end{equation}
has another advantage compared to manifolds in Figures~\ref{oldfig2} and \ref
{rep_man_for_log}. Unlike those, for $\lambda \gg 1$ and small enough $%
p-p_{0}$, manifold (\ref{MAN3}) has all caustics in a safe distance. The
RMs are horizontal lines $p_{n}=p_{0}+n\hbar \alpha $,
independent of $x$ and corresponding transformation elements can be evaluated exactly as 
$\langle p|p_{0}\rangle _{n}=\delta (p-p_{0}-n\hbar \alpha )$. Using (\ref
{RM_SUM}), the uniform expression for $\langle p|p_{0}\rangle $ is 
\begin{equation}
\langle p|p_{0}\rangle _{unif}=\sum_{n=-\infty }^{\infty }J_{n}\left(\frac{%
\lambda }{2}\right)\,i^{n}\,\delta (p-p_{0}-n\hbar \alpha ),  \label{UNIF_PROP3}
\end{equation}
so the wave function is determined by its Fourier coefficients $%
c_{n}(\lambda )=J_{n}(\lambda /2)\,i^{n}$, which is natural since manifold (%
\ref{MAN3}) is exactly periodic. Moreover, each RM
contributes only to a single momentum ($p_{n}$). Put differently, we have
calculated the wave function for any specific momentum to all orders in $%
\lambda $. Now let us find the semiclassical form of $\langle p|p_{0}\rangle 
$.
For a given momentum $p$ there is an infinite number of SP
points which occur in pairs $(x_{n},y_{n})$, 
\begin{equation}
x_{n}=x_{0}+\frac{2\pi }{\alpha }n,\ y_{n}=\frac{\pi }{\alpha }-x_{0}+\frac{%
2\pi }{\alpha }n,  \label{SP_POINTS_3}
\end{equation}
where $x_{0}=-\frac{1}{\alpha }\arcsin \frac{2(p-p_{0})}{\lambda \hbar
\alpha }$. Analogously to (\ref{SP_INT}), the semiclassical transformation
element for manifold (\ref{MAN3}) is 
\begin{eqnarray}
\langle p|p_{0}\rangle _{sc} &=&\frac{1}{\sqrt{2\pi \hbar }}\sum_{n=-\infty
}^{\infty }\left| \frac{\partial ^{2}S}{\partial x^{2}}\right|
_{x=x_{n}}^{-1/2}  \label{SC_PROP3_1} \\
&\times &\left\{ \exp \left[ \frac{i}{\hbar }S(x_{n})-i\frac{\pi }{4}\right]
+\exp \left[ \frac{i}{\hbar }S(y_{n})+i\frac{\pi }{4}\right] \right\}  
\nonumber
\end{eqnarray}
with
\[
S(x,p,p_{0})=-px+F_{2}(x,p_{0})=(p_{0}-p)\,x+\frac{1}{2}\hbar \lambda
\cos \alpha x.
\]
Using (\ref{SP_POINTS_3}), (\ref{SC_PROP3_1}) and defining 
$r=\frac{p-p_{0}}{\hbar \alpha }$, we get
\begin{equation}
\langle p|p_{0}\rangle _{sc}=\frac{1}{\hbar \alpha }F(\lambda
,r)\sum_{n=-\infty }^{\infty }e^{2\pi inr}
\end{equation}
where 
\begin{eqnarray}
F(\lambda ,r)&&=\sqrt{\frac{2}{\pi }}e^{-ir\pi /2} \\
\times &&\frac{\cos \left\{ r\left( \frac{\pi }{2}+\arcsin \frac{2r}{%
\lambda }\right) +\left[ \left( \frac{\lambda }{2}\right) ^{2}-r^{2}\right]
^{1/2}-\frac{\pi }{4}\right\} }{\left[ (\frac{\lambda }{2})^{2}-r^{2}\right]
^{1/4}}.  \nonumber
\end{eqnarray}
Employing the Poisson summation formula 
\[
\sum_{n=-\infty }^{\infty }e^{2\pi
inr}=\sum_{m=-\infty }^{\infty }\delta (m-r)
\]
and reverting to $%
p=p_{0}+r\hbar \alpha $, we obtain 
\begin{eqnarray}
\langle p|p_{0}\rangle _{sc} &=&\frac{1}{\hbar \alpha }F(\lambda
,r)\sum_{n=-\infty }^{\infty }\delta (n-r)  \label{SC_PROP3_2 } \\
&=&\frac{1}{\hbar \alpha }\sum_{n=-\infty }^{\infty }F(\lambda ,n)\,\delta
(n-r)  \nonumber \\
&=&\sum_{n=-\infty }^{\infty }F(\lambda ,n)\,\delta (P-p_{0}-n\hbar \alpha ).
\nonumber
\end{eqnarray}
Comparing (\ref{SC_PROP3_2 }) with (\ref{UNIF_PROP3}), we see that the
semiclassical and uniform versions of $\langle p|p_{0}\rangle $ will be
asymptotically equal if $F(\lambda ,n)\sim J_{n}(\lambda /2)\,i^{n}$ for
large $\lambda $. That is indeed true since for $\lambda \gg n^{2}$, 
\begin{eqnarray}
F(\lambda ,n) &\sim &\left( \frac{2}{\pi \frac{\lambda }{2}}\right)
^{1/2}e^{in\pi /2}\cos \left( \frac{\lambda }{2}-\frac{n\pi }{2}-\frac{\pi }{%
4}\right)  \\
&\sim &J_{n}\left( \frac{\lambda }{2}\right) i^{n}  \nonumber
\end{eqnarray}
Table \ref{tbl} shows that for $\lambda =10$ semiclassical and
RM values of $a_{n}$ differ by less than $0.02$ up to $n=3$%
. In the opposite limit, for $\lambda <1$, even $a_{0}$ evaluated
semiclassically is completely off.
\begin{table}[tbp]
\caption{ Comparison of numerical, RM, and
SP evaluation of $|\phi (p)|^{2}$ for $\xi (x)=\alpha x$, $%
\lambda =10$. }
\label{tbl}
\begin{tabular}{|l|llllll|}
$n$ & 0 & 1 & 2 & 3 & 4 & 5 \\ \hline
$|a_n|$ (num.) & 0.1776 & 0.3276 & 0.0466 & 0.3648 & 0.3912 & 0.2611 \\ 
\hline
$|a_n|$ (RM) & 0.1776 & 0.3276 & 0.0466 & 0.3648 & 0.3912 & 0.2611 \\ \hline
$|a_n|$ (SP) & 0.1704 & 0.3324 & 0.0343 & 0.3622 & 0.4312 & $\infty$%
\end{tabular}
\end{table}

\section{Scattering from a corrugated wall}

\label{WALL}

\begin{figure}[htbp]

\centerline{\epsfig{figure=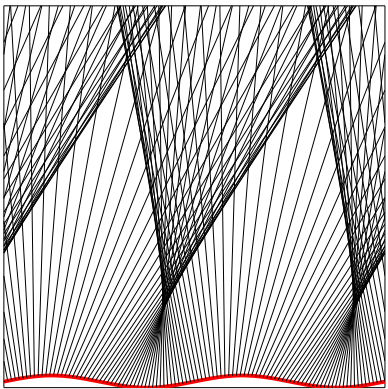,width=\myhsize}}
\vspace{\myvspace}
\caption{Ray picture for a two-dimensional scattering from a corrugated
wall. Only reflected rays are shown.}
\label{ray_picture}

\end{figure}

\begin{figure}[htbp]

\centerline{\epsfig{figure=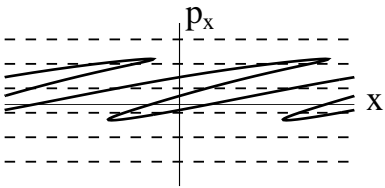,width=\myhsize}}
\vspace{\myvspace}
\caption{Original (solid line) and replacement manifolds (dashed lines) on
the Poincar\'{e} surface of section for scattering from a corrugated wall.}
\label{poincare_surface}

\end{figure}

Let us apply the replacement-manifold method to a physical problem
generating loop structure in phase space. Scattering of a plane wave by a
corrugated wall has exactly such a property. This system has been used to model elastic scattering of atoms by solid surfaces (for review see 
\cite{hoinkes}). Consider a two-dimensional coordinate space divided into
two parts by a periodically curved boundary consisting of the set of points
with coordinates related by 
\begin{equation}
\tilde{y}(x)=\frac{\varepsilon }{\beta }\sin \beta x  \label{BOUNDARY}
\end{equation}
where $\beta $ gives the spatial frequency of corrugation and $\varepsilon $
the maximum slope $\frac{d\tilde{y}}{dx}$ of the wall. A plane wave with
momentum $\hbar {\bf k}_{1}$ is incident from the upper half-plane at an
angle $\alpha _{1}$ from the $y$ axis, so the incident wave vector is ${\bf k%
}_{1}=k\,(\sin \alpha _{1},-\cos \alpha _{1}).$ Classically, the wave
reflects specularly from the curved boundary. The reflected rays are shown
in Fig.~\ref{ray_picture}.
 Notice also the clearly visible caustics. The
Poincar\'{e} surface of section in Fig.~\ref{poincare_surface} displays
dependence of momentum component $p_{2x}$ of reflected rays on coordinate $x$
at a given distance $y_{2}$ from $x$ axis. We can see familiar loops with
constant area and predict the failure of semiclassical approximation when
this area gets smaller than $\hbar $. Below we uniformize the semiclassical
solution using RMs and compare the outcome with the exact
and primitive semiclassical results. 

\begin{figure}[htbp]

\centerline{\epsfig{figure=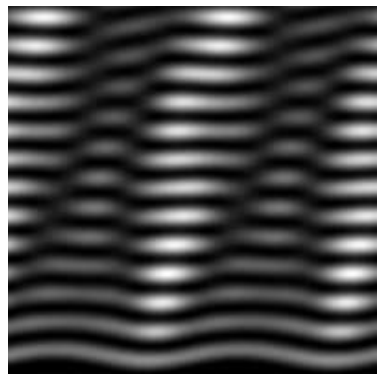,width=\myhsize}}
\vspace{\myvspace}
\caption{Probability density plot for the exact quantum solution. Plot
  of the RM solution is indistinguishable to the eye. For detailed
  comparison, see Fig.~\ref{wf_on_section}. }
\label{quantumWF}

\end{figure}

There exist several approaches to solve this scattering problem exactly.
Garcia and Cabrera \cite{garcia} have thoroughly compared  merits of various
methods. The main issue is the solution's convergence, which improves with
decreasing corrugation parameter $\epsilon $. Luckily, this regime overlaps
with the small-loop limit in which we are interested. We can therefore use the most
straightforward approach based on the so-called Rayleigh hypothesis
 \cite{rayleigh} which has been shown to be valid for $\epsilon <0.448$ by Millar 
\cite{millar}. Details of the solution are provided in the Appendix. A
probability density plot of the wave function is shown in Fig.~\ref{quantumWF}.

The whole scattering problem can be formulated using an analogue to the
smooth-potential Lippmann-Schwinger equation. Invoking the Green's theorem,
it can be shown that the total wave function satisfies 
\begin{equation}
|\psi _{tot}\rangle =|\psi _{inc}\rangle
+\hat{G}_{0}\int_{boundary}\!\!\!\!\!\!\!\!\!\!\!dx\,|{\bf r}\rangle \,\hat{n}%
(x)\cdot {\bf \nabla }\psi _{tot}({\bf r)}  \label{LIPPMANN}
\end{equation}
where $\hat{G}_{0}$ is the free-space Green's operator and $\hat{n}(x)$ is a
normal unit vector at the boundary pointing into free space. Now we could
use periodicity of the wall, expand $\psi _{tot}({\bf r})$ in its Fourier
modes,
\begin{eqnarray}
\psi _{tot}({\bf r}) 
&=&
\psi _{inc}({\bf r}) + \psi _{scat}({\bf r}) \nonumber \\
&=& 
\frac{1}{2\pi } e^{i{\bf k}_{1}\cdot {\bf r}}
+ \frac{1}{2\pi }
\sum_{n=-\infty }^{\infty }a_{n}e^{i{\bf k}_{2,n}\cdot {\bf r}},
\label{EXPANSION}
\end{eqnarray}
 and recover the linear system (\ref{BC_4}) for coefficients $a_{n}$.
Our goal, however, is to make a connection with a semiclassical picture in
the Poincar\'{e} surface of section (Fig.~\ref{poincare_surface}). That
figure implicitly assumes that classically only a single scattering event
takes place before a ray leaves the wall permanently. If corrugation is deep
enough ($\varepsilon >\varepsilon _{\max }$, where $\varepsilon _{\max }<1$%
), multiple scattering will occur for any incident angle $\alpha _{1}$. The
larger $\varepsilon $ gets (within the range $0<\varepsilon <\varepsilon
_{\max }$), the smaller the maximum allowed incident angle $\alpha _{1}$.
Let us consider a wall with very shallow corrugation ($\varepsilon =0.2)$
and a small angle of incidence ($\alpha _{1}=0.2)$, a situation which
classically  allows  only a single reflection. We also choose the
incident-wave length $\lambda =\frac{2\pi }{k}$ to be $0.3$ times the period
of corrugation. The surprising result we shall obtain below is that although
there exist seven real Bragg peaks in the scattered wave, the semiclassics
break down while the single-scattering approximation using replacement
manifolds works with excellent accuracy. 

\begin{figure}[htbp]

\centerline{\epsfig{figure=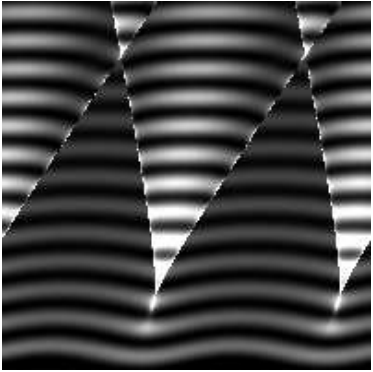,width=\myhsize}}
\vspace{\myvspace}
\caption{Probability density plot for the semiclassical solution.}
\label{semiclassicalWF}

\end{figure}

\begin{figure}[htbp]

\centerline{\epsfig{figure=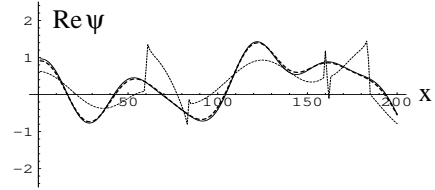,width=\myhSize}}
\vspace{\myvspace}
\caption{Wave function on the surface of section at $y = 4 \pi / \beta$. Comparison of the
exact quantum (dashed line), RM (solid line), and semiclassical (dotted
line) wave functions.}
\label{wf_on_section}

\end{figure}

The simple semiclassical wave
function may be evaluated e.g. by tracing of individual rays and employing
the Van Vleck propagator~\cite{vanVleck}. The action must be adjusted by correct Maslov
indices, corresponding to reflection from a hard wall and to passage through
caustics~\cite{gutzwiller}. A probability density plot of the semiclassical wave function is shown in
Fig.~\ref{semiclassicalWF}. Note the caustics, which clearly separate regions with one,
three, and five contributions to the scattered wave. Between caustics,
solution looks qualitatively the same as the exact quantum analogue in Fig.~\ref{quantumWF}. By looking at a surface of section at $y_{2} = 4 \pi / \beta$ (where there are already at least three contributions for any 
$x$) in Fig.~\ref{poincare_surface}, we expect that semiclassical and quantum
wave functions should disagree everywhere. This guess is confirmed in Fig.~\ref{wf_on_section}.

Now let us find a uniform expression for the scattered wave. The
Lippmann-Schwinger equation (\ref{LIPPMANN}) is an integral equation, and it
is not immediately obvious how it could be used to find $\psi _{scat}$ from $%
\psi _{inc}$. We circumvent this problem by proceeding in four steps: first
we canonically transform the incident wave function to a new coordinate
system ${\bf r}^{\prime }$ in which the wall becomes straight. We solve the
scattering problem in these coordinates, since there the semiclassical
approach is exact. Then we transform back to the original coordinate system
in which we propagate the wave using the free-space Green's function to
obtain our final answer. Semiclassical approach is used for each separate
step but their combination is evaluated exactly.

The wall becomes flat if we apply a canonical transformation 
\[
x^{\prime }=x,\quad y^{\prime }=y-\frac{\varepsilon }{\beta }\sin \beta x
\]
generated by 
\begin{equation}
F_{2}({\bf r}^{\prime },{\bf k})=k_{x}x^{\prime }+k_{y}(y_{{}}^{\prime }+%
\frac{\varepsilon }{\beta }\sin \beta x^{\prime }).
\end{equation}
The incident wave function $\psi _{inc}({\bf k}_{1})$ can be transformed
into new coordinates using 
\begin{equation}
\langle {\bf r}^{\prime }|{\bf k}\rangle _{sc}=\frac{1}{2\pi }\,e^{iF_{2}(%
{\bf r}^{\prime },{\bf k})}  \label{CAN_TRANS1}
\end{equation}
In primed coordinates, the semiclassical solution of scattering is exact and
``angle of incidence'' equals ``angle of reflection''. Therefore $\hat{n}%
^{\prime }\cdot {\bf \nabla }\psi _{tot}({\bf r}^{\prime })$ may be replaced
by $2\hat{n}^{\prime }\cdot {\bf \nabla }\psi _{inc}({\bf r}^{\prime })$ in
equation (\ref{LIPPMANN}). The non-diagonal part of the semiclassical
scattering matrix becomes 
\begin{equation}
\hat{T}=\hat{S}-\hat{1}=2\left. \int dx^{\prime }\,|{\bf r}^{\prime }\rangle
\,\hat{n}^{\prime }\cdot {\bf \nabla }^{\prime }\langle {\bf r}^{\prime
}|\right| _{y^{\prime }=0}  \label{TRANSFER}
\end{equation}
where $\hat{n}^{\prime }=\hat{y}^{\prime }$. Transformation back to original
coordinates is achieved by a complex conjugate of (\ref{CAN_TRANS1}). The
free-space propagation is accomplished by a semiclassical Green's function,
exact in mixed representation 
\begin{equation}
\langle k_{2x},y_{2}|\hat{G}_{0}^{sc}|{\bf k}\rangle =\frac{\delta
(k_{2x}-k_{x})\,e^{ik_{y}y_{2}}}{\sqrt{2\pi }(k^{2}-k_{2}^{2}+i\epsilon )}
\label{FREE_SPACE}
\end{equation}
Putting all four pieces together, we obtain an expression for the scattered
wave on a surface of section, 
\begin{eqnarray}
&\langle &k_{2x},y_{2}|\hat{G}_{0}^{sc}\hat{T}\,|{\bf k}_{1}\rangle
\label{ALL_PIECES1} \\
&=&2\int d^{2}k\left. \int dx^{\prime }\langle k_{2x},y_{2}|\hat{G}_{0}^{sc}|%
{\bf k}\rangle \langle {\bf k}|{\bf r}^{\prime }\rangle \,\partial
_{y^{\prime }}\langle {\bf r}^{\prime }|{\bf k}\rangle _{sc}\right|
_{y^{\prime }=0}  \nonumber   \\
&=&\frac{2ik_{1y}}{(2\pi )^{5/2}}\int d_{{}}^{2}k\,\frac{\delta
(k_{2x}-k_{x})\,e^{ik_{y}y_{2}}}{\sqrt{2\pi }(k^{2}-k_{2}^{2}+i\epsilon )} \nonumber \\
&\times &\int dx^{\prime }\,\exp [i(k_{1x}-k_{x})x^{\prime }+i(k_{1y}-k_{y})%
\frac{\varepsilon }{\beta }\sin \beta x^{\prime }]  \nonumber
\end{eqnarray}

Before we proceed, it should be noted that the outcome of this
four-step process, an 
integral representation of the scattered wave, can be viewed as a
continuous superposition of waves emanating from sources along the
boundary, with strength proportional to the normal derivative of the
incident wave. It can be shown, in fact, that this process is
equivalent to a generalized Kirchhoff diffraction method \cite{kirchhoff}.

Returning to expression~(\ref{ALL_PIECES1}), we recognize that the
exponent in integral over $x^{\prime }$  corresponds to a classical manifold  
\begin{equation}
k_{x}=k_{1x}+(k_{1y}-k_{y})\varepsilon \cos \beta x^{\prime }
\end{equation}
which is ``begging'' to be replaced by partial manifolds $%
k_{x,n}=k_{1x}+n\beta x^{\prime }$ with weights $J_{n}(\lambda )$ because 
\[
|\lambda |=|k_{1y}-k_{y}|\frac{\varepsilon }{\beta }\lesssim \frac{%
2k\varepsilon }{\beta }<1
\]
for the classically allowed momenta. The integral
over $x^{\prime }$ is then simple to evaluate and is equal to 
\begin{eqnarray}
\int dx^{\prime } &&\sum_{n=-\infty }^{\infty }J_{n}\left( \lambda \right)
\exp [i(k_{1x}+n\beta -k_{x})\,x^{\prime }] \\
&=&2\pi \sum_{n=-\infty }^{\infty }J_{n}(\lambda )\,\delta (k_{1x}+n\beta
-k_{x})  \nonumber
\end{eqnarray}
Using this result and evaluating the trivial integral over $k_{x}$,
expression (\ref{ALL_PIECES1}) becomes 
\begin{eqnarray}
&\langle &k_{2x},y_{2}|\hat{G}_{0}^{sc}\hat{T}\,|{\bf k}_{1}\rangle =\frac{%
2ik_{1y}}{(2\pi )^{3/2}} \\
&\times &\sum_{n=-\infty }^{\infty }\int dk_{y}\,\frac{\,e^{ik_{y}y_{2}}}{\sqrt{%
2\pi }(k^{2}-k_{2}^{2}+i\epsilon )}\,J_{n}(\lambda )\,\delta
(k_{2x}-k_{2x,n})  \nonumber
\end{eqnarray}
where $k_{2x,n}=k_{1x}+n\beta $. Integral over $k_{y}$ picks up a pole at $%
k_{y}=k_{2y,n}=\sqrt{k^{2}-k_{2x,n}^{2}}$ and the final answer is 
\begin{eqnarray}
\langle k_{2x},y_{2} &|&\hat{G}_{0}^{sc}\hat{T}\,|{\bf k}_{1}\rangle  \\
&=&\frac{k_{1y}}{\sqrt{2\pi }}\sum_{n=-\infty }^{\infty }e^{ik_{2y,n}y_{2}}\,%
\frac{J_{n}(\lambda _{n})\,\delta (k_{2x}-k_{2x,n})}{k_{2y,n}}  \nonumber
\end{eqnarray}
where $\lambda _{n}=(k_{1y}-k_{2y,n})\frac{\varepsilon }{\beta }$. The scattered wave is thus a superposition of traveling and evanescent waves with wavevectors ${\bf k}_{2,n}$ with coefficients 
\begin{equation}
a_{n,unif}=\frac{k_{1y}}{k_{2y,n}}J_{n}(\lambda _{n}).
\label{UNIF_COEFF}
\end{equation}
These turn out to be very close to the exact quantum coefficients obtained in the Appendix, but here we have avoided the solution of a linear system (\ref{BC_4}).  

The uniform
expression (\ref{UNIF_COEFF}) for coefficients of expansion
(\ref{EXPANSION}) is surprisingly accurate. A two-dimensional density plot in Fig.~\ref{quantumWF} does not reveal any difference from the exact quantum solution
while an analogous semiclassical plot clearly shows the caustics (see
Fig.~\ref{semiclassicalWF}). Even if we look at the Poincar\'{e} surface of section in
a region with many classical caustics, the quantum and uniform solutions
agree while semiclassical solution fails miserably (see Fig.~\ref{wf_on_section}).
 The uniform solution agrees with one found by
Garibaldi et al.\cite{garibaldi} and is somewhat more accurate than
the same result without the $k_{1y} / k_{2y,n}$ prefactor obtained
by Hubbard and Miller \cite{hubbard} by SCP approximation and by Garibaldi et al. \cite{garibaldi}. 
As Garibaldi et al. (who obtain three different solutions differing only by prefactors) point out, these prefactors ``should not be taken too seriously.'' We agree:  for traveling modes, they do not cause large errors, and since all these solutions neglect multiple scattering, we cannot expect high accuracy of the already small coefficients of evanescent modes.    

As promised in Section \ref{GENERAL}, we have obtained almost identical result using RMs as Hubbard and Miller with SCP approximation \cite{hubbard}. They found their answer for a hard wall as a limit of the solution for a smooth potential due to Lennard-Jones and Devonshire\cite{lennard}.
Although we were lucky in this case to obtain
exactly solvable integrals for RMs, we can quite
generally expect that the replacement manifolds will be much smoother than
the original one, enabling their simple evaluation by SP method. 

We discuss the specific choice of parameters. Selecting $\varepsilon \ll 1$, 
$\alpha \ll 1$ was necessary for the validity of the single-scattering
approximation. At first, it appears that the simple semiclassical approximation should be
accurate for this regime of very shallow corrugation, since scattering from
a flat wall has an exact semiclassical answer. The correct criterion,
however, is based on the area of loops in phase space (see Fig.~\ref{poincare_surface}).
In this case, for small $\varepsilon $ and $\alpha $, the height of loops
is equal to the momentum kick in region of maximum slope, approximately $%
p\varepsilon $ and the width is half the period of corrugation, $\frac{\pi }{%
\beta }$. The loop area is smaller than $p\varepsilon \frac{\pi }{\beta }=$ $%
\pi kd\hbar $ ($d$ is the depth of corrugation) and has to be compared to $\hbar 
$. We arrive at a surprising result that the validity of semiclassical
approximation has nothing to do with the periodicity of corrugation, but
only depends on the product $kd$ of wave vector and corrugation depth. If
this product is small, semiclassical approximation breaks down. In the
opposite case of deep corrugation, semiclassical approximation works,
although we might have to take into account classical trajectories
corresponding to multiple scattering. 

\section{Conclusion}

We have successfully uniformized several intimidating situations which can
arise when repeated small areas or loops arise from enclosing classical
manifolds using the replacement-manifold idea. This notion can rejuvenate
physical intuition about the quantum wave function and also be a very
convenient approach to an accurate result. Two physically motivated cases
were considered (the homoclinic-like oscillation and the corrugated wall)
along with several other contrived cases to test the method. The present
method should work in more complicated time-dependent problems if used with
Miller's IVR applied to Van Vleck's semiclassical propagator. After
replacement manifolds are identified, the perturbative expansion
contributions to the wave function could be found by SP approximation
applied to these manifolds.

Future work along these lines includes the attempt to uniformize  a full
homoclinic tangle.  It remains to be seen how convenient and generally
applicable the replacement manifold idea is.  However, it seems clear
that we now have a new method which simplifies many problems while
reinstating an intuitive foundation.

\section{Acknowledgements}

This research was supported by the National Science Foundation under Grant
No. CHE9610501 and by the Institute for Theoretical Atomic and Molecular
Physics. One of us (J. V.) would like to acknowledge helpful discussions
with A. Mody and A. Barnett.

\appendix

\section{Exact solution of scattering from a hard sinusoidal wall}

From conservation of energy, the scattered and incident waves must have the
same magnitude of momentum. Due to the periodicity of the wall in the
$x$ direction, the $x$-component of
momentum  is constrained by the Bragg condition 
\begin{equation}
k_{2x,n}=k_{1x}+n\beta ,\;\;\;\;n\in Z.  \label{K2XN}
\end{equation}
Corresponding $y$-component of momentum $k_{2y,n}=\sqrt{k^{2}-k_{2x,n}^{2}}$
will be real only if $|k_{1x}+n\beta |<k$. Otherwise the reflected wave
becomes evanescent: it decays exponentially with distance from $x$-axis
because $k_{2y}$ is imaginary. To complete the solution, coefficients $a_{n}$
in the momentum Green's function 
\[
\langle {\bf k}_{2x}|{\bf k}_{1x}\rangle =\sum_{n=-\infty }^{\infty
}a_{n}\,\delta (k_{2x}-k_{2x,n})\,\delta (k_{2y}-k_{2y,n}) 
\]
must be found. This is accomplished by imposing the Dirichlet boundary
condition in coordinate space. Namely, the total wave function $\psi _{tot}(%
{\bf r})=\psi _{inc}({\bf r})+\psi _{scat}({\bf r})$ has to vanish along the
corrugated wall. Using 
\begin{eqnarray}
\psi _{inc}({\bf r})&=&\langle {\bf r}|{\bf k}_{1}\rangle =\frac{1}{2\pi }%
e^{i{\bf k}_{1}\cdot {\bf r}}, \\
\psi _{scat}({\bf r}) &=& \int d^{2}k_{2}\,\langle {\bf r}|{\bf k}%
_{2}\rangle \langle {\bf k}_{2}|{\bf k}_{1}\rangle = \frac{1}{2\pi }%
\sum_{n=-\infty }^{\infty }a_{n}e^{i{\bf k}_{2,n}\cdot {\bf r}},  \nonumber
\end{eqnarray}
definition of boundary (\ref{BOUNDARY}) and (\ref{K2XN}) for $k_{2x,n}$, the
boundary condition becomes 
\begin{eqnarray}
&& \exp \left[ i(k_{1x}x+k_{1y}\frac{\varepsilon }{\beta }\sin \beta x) \right] \\
&+& \sum_{n=-\infty
}^{\infty }a_{n} \exp \left[ i(k_{1x}+n\beta )x+ik_{2y,n}\frac{\varepsilon }{\beta }%
\sin \beta x \right] = 0. \nonumber
\end{eqnarray}
Recalling the identity $e^{i\lambda \sin \xi (x)}=\sum_{m=-\infty }^{\infty
}J_{m}(\lambda )\,e^{imx}$ and denoting $\lambda _{1}=\frac{\varepsilon }{%
\beta }k_{1y}$, $\lambda _{2,n}=\frac{\varepsilon }{\beta }k_{2y,n}$, $%
\theta =\beta x$, the boundary condition may be written as 
\begin{equation}
\sum_{m=-\infty }^{\infty }\left[ J_{m}(\lambda _{1})\,e^{im\theta
}+\sum_{n=-\infty }^{\infty }a_{n}\,J_{m}(\lambda _{2,n})\,e^{i(n+m)\theta
}\right] =0.  \label{BC_2}
\end{equation}
Replacing $m$ by $p\equiv n+m$ in the second term and by $p\equiv m$ in the
first term in the brackets in (\ref{BC_2}), we get 
\begin{equation}
\sum_{p=-\infty }^{\infty }e^{ip\theta }\left[ J_{p}(\lambda
_{1})\,+\sum_{n=-\infty }^{\infty }a_{n}\,J_{p-n}(\lambda _{2,n})\,\right]
=0.
\end{equation}
This equation will be satisfied for all $x=\frac{\theta }{\beta }$ only if
coefficients of $e^{ip\theta }$ on the left-hand side vanish, 
\begin{equation}
J_{p}(\lambda _{1})\,+\sum_{n=-\infty }^{\infty }a_{n}\,J_{p-n}(\lambda
_{2,n})=0,\;\;\;\;\forall p\in Z.  \label{BC_4}
\end{equation}
We recognize (\ref{BC_4}) as an infinite system of linear equations for
coefficients $a_{n}$. In matrix form, ${\bf A\cdot a=b}$, with $A_{ij}\equiv
J_{i-j}\left( \frac{\varepsilon }{\beta }\sqrt{k^{2}-(k_{1x}+j\beta )^{2}}%
\right) $and $b_{i}\equiv -J_{i}(\frac{\varepsilon }{\beta }k_{1y})$.
Although this system cannot be solved analytically, expressing the solution
in this form is a great improvement compared to what can be done for a
general Dirichlet problem satisfying time-independent Schr\"{o}dinger
equation. The system (\ref{BC_4}) can be approximately evaluated numerically
by restricting $n$ and $p$ to lie between limiting values $\pm n_{\max }$.


\begin{references}

\bibitem{softChaos}  Surprisingly enough, softly chaotic systems
resist semiclassical treatment often more than those with strongly chaotic behavior. It is
because loop areas in phase space increase as chaos grows stronger.
For semiclassical analysis of a strongly chaotic system see e.g. S. Tomsovic and
E. J. Heller, Phys. Rev. E {\bf \ 47}, 282 (1993) or P. W. O'Connor, S.
Tomsovic, E. J. Heller, J. Stat. Phys. {\bf 68}, 131 (1991). For
semiclassical analysis of a mixed system (with weak chaos) see e.g. M. A. Sep\'{u}%
lveda and E. J. Heller, J. Chem. Phys. {\bf 101}, 8004 (1994).

\bibitem{ivr}  W. H. Miller: J. Chem. Phys., {\bf 53}, 3578 (1970).

\bibitem{millerIVR}  W. H. Miller: Adv. Chem. Phys., {\bf 25}, 69
(1974).

\bibitem{rick}  E. J. Heller, ``The Role of Interference in the
Semiclassical Approximation to Chaotic Motion'', {\it Prog. Theor. Physics
Supplement}, {\bf N116}, 45 (1994).

\bibitem{airyCarrier} G. F. Carrier. J. Fluid Mech. {\bf 24}, 641 (1966).

\bibitem{airyBerry} M. V. Berry, Proc. Phys. Soc. (London) {\bf 89}, 479 (1966).

\bibitem{canonical}  We keep the discussion in terms of general canonical
transformations although we often have in mind the semiclassical Van Vleck
propagator when speaking of transformation elements.

\bibitem{airyUnif}  Even if the integral is not analytic, an Airy-function
uniformization procedure can be used. See, e.g. \cite{airyCarrier,airyBerry}.

\bibitem{millerSPT}  W. H. Miller and F. T. Smith, Phys. Rev. A {\bf 17},
939 (1978).

\bibitem{hubbard}  L. M. Hubbard and W. H. Miller, J. Chem. Phys. {\bf \ 78}%
, 1801 (1983).

\bibitem{complexSP}  The small difference, especially between the two contributions, is the result of our neglecting the shift of SP points due to the finite extent of $\phi ^{i}(p_{0})$. 

\bibitem{hoinkes}  H. Hoinkes, Rev. Mod. Phys. {\bf 52}, 933 (1980).

\bibitem{garcia}  N. Garcia and N. Cabrera, Phys. Rev. B {\bf 18}, 576
(1978).

\bibitem{rayleigh}  J. W. Strutt (Baron Rayleigh), Proc. R. Soc. Lond. A 
{\bf 79}, 399 (1907).

\bibitem{millar}  R. F. Millar, Proc. Camb. Philos. Soc. {\bf 69}, 217
(1971).

\bibitem{vanVleck} J. H. Van Vleck, Proc. Natl. Acad. Sci. {\bf 14}, 178 (1928).

\bibitem{gutzwiller} M. C. Gutzwiller, {\it Chaos in Classical and Quantum Mechanics} (Springer, Berlin, 1990).

\bibitem{kirchhoff}
The Kirchhoff diffraction method starts from the Weber integral
representation (analogue of a more familiar Helmholtz representation
in three dimensions)
\[
\psi ({\bf r}_2) = \int_{C} ds \left( \psi \frac{\partial}{\partial n}
  H^{(0)}_1(kR) - H^{(0)}_1(kR) \frac{\partial \psi}{\partial n}
   \right),
\]
where $R=|{\bf r}_2 - {\bf r}|$, ${\bf r}$ is the coordinate vector
of points on the closed curve $C$, $\frac{\partial}{\partial n}$ is
the derivative in the direction of an {\it outward} pointing normal vector
$\hat{\bf n}$.  For derivation of Weber and Helmholtz
representations, see e.g. B. B. Baker and E. T. Copson, The Mathematical Theory
of Huygens' Principle, 2nd ed., Oxford University Press, London, 48 (1950) and
M. Born and E. Wolf, Principles of Optics, 6th ed., Pergamon Press,
Elmsford, 375 (1980).
 
Kirchhoff method consists in approximating usually unknown values of
$\psi$ and $\frac{\partial \psi}{\partial n}$ in the integrand.  Originally, it
was only used in the case of a plane screen with an aperture: Kirchhoff
set $\psi$ and $\frac{\partial \psi}{\partial n}$ equal to zero on the
dark side of the screen and equal to the unperturbed values
($\psi_{inc}$ and $\frac{\partial \psi_{inc}}{\partial n}$) in the
aperture.  The term ``Kirchhoff method'' is now sometimes used for a
more general approximation encompassing our example of corrugated
wall, whereby $\psi$ and $\frac{\partial \psi}{\partial n}$ at any
point of the surface are
approximated by the values that would be present on the tangent plane
at that point.  See, e.g., P. Beckmann and  A. Spizzichino, The
Scattering of Electromagnetic Waves from Rough Surfaces, Artech House,
Norwood, 20 (1987). 

\bibitem{garibaldi}  U. Garibaldi, A. C. Levi, R. Spaldacini and G. E.
Tommei, Surface Science {\bf 48}, 649 (1975).

\bibitem{lennard} J. E. Lennard-Jones and A. F. Devonshire, Proc. R. Soc. London Ser. A {\bf 158}, 242 (1937).

\end{references}
\end{document}